\documentclass[aps,prd,twocolumn,showpacs,preprintnumbers,amsmath,amssymb,nofootinbib]{revtex4-1}
\usepackage{amsmath,amsfonts,euscript,amssymb}
\usepackage{epsfig}
\DeclareSymbolFont{bfitletters}{OML}{cmm}{bx}{it}
\DeclareSymbolFont{bfitoperators}   {OT1}{cmr} {m}{n}
\DeclareMathSymbol{\bfitomega}{\mathord}{bfitletters}{"21}
\DeclareMathSymbol{\bfitrho}{\mathord}{bfitletters}{"1A}
\DeclareMathSymbol{\bfitLambda}{\mathord}{operators}{"03}
\DeclareMathSymbol{\lambda}{\mathord}{letters}{"15}
\DeclareMathSymbol{\bfitgamma}{\mathord}{bfitletters}{"0D}

\newcommand{\be}{\begin{equation}}
\newcommand{\ee}{\end{equation}}
\newcommand{\bea}{\begin{eqnarray}}
\newcommand{\eea}{\end{eqnarray}}
\begin{document}

\title{Cycles of Time in Classical Cosmology}

\author{\bf A. E. Pavlov}
\affiliation{Institute of Mechanics and Power Engineering, Russian State Agrarian University -- Moscow Timiryazev Agricultural Academy, Timiryazevskaya str. 49,
Moscow 127550, Russia\\
alexpavlov60@mail.ru}

\begin{abstract}
We present exact solutions to the Friedmann equation in standard ${\rm \Lambda CDM}$ cosmology in Weierstrass and Jacobi functions. The right hand side of the Friedmann equation describing various contributions of matter sources is considered in generic form. It is proved that the problem of integration of the Friedmann equation for simple EoS of medium is reduced to solving Abel integrals for algebraic functions.
\end{abstract}

\pacs{04.20.Cv, 98.80.Es}

\maketitle

\section{Introduction}

The last few decades have seen a remarkable progress in cosmology~\cite{RiessNobel}. Observations with more advanced instruments will resolve mysterious questions in the coming decades. The Friedmann equation is used to interpret the observational data. It connects the rate of expansion of the Universe with the energy density of matter and spatial curvature. For different states of matter, different scenarios of the evolution of the Universe are obtained
~\cite{Behnke,Zakharov:2010nf,PavlovMIPh}.

Cosmologists prefer the numerical integration of the Friedmann equation. Since the right hand side of the Friedmann equation has a polynomial form, it is of interest to use the theory of doubly periodic functions to find analytical solutions developed in papers of Jacobi, Abel, Weierstrass, Kovalevskaya. The theory has found effective application in problems of analytical mechanics~\cite{Golubev} and celestial mechanics~\cite{Gerasimov}.

Kovalevskaya raised the problem of finding all integrable cases of rigid body rotations in class of meromorphic at all plane of complex time variable. This statement of the question presented essential extension of the original mechanical problem. Moreover, this extension had purely mathematical character and was not demanded by any mechanical argumentations. This remarkable mathematical extension of studying the mechanical problem on the plane of complex time permitted applying the theory of analytical functions excellent elaborated in ${\rm XIX}$-th century.
Later on, successfully, the approach was taken into consideration by mathematicians in studying various problems in applied mechanics.

We get more information about real functions considering them on a complex plane~\cite{PenroseRus}.
The application of meromorphic functions of complex time in theoretical cosmology is of interest. In this work, by including the most significant contributions from different states of matter, we obtain analytical solutions of the Friedmann equation. The author hopes the analytical solutions will be useful in the interpretation of the modern cosmological data.

\section{Conformal Friedmann equation}

In the standard ${\rm \Lambda CDM}$ cosmology the conformal Friedmann equation~\cite{Weinberg} is represented by the first integral
\begin{equation}\label{Friedmann}
\left(\frac{a'}{a}\right)^2=\frac{8\pi G}{3}a^2\rho-k.
\end{equation}
Here $a(\eta)$ is the scale factor, the prime denotes the derivative with respect to the conformal time $\eta$, $G$ is the Newton constant, $\rho$ is the  matter sources density, $k$ is the constant of space curvature. The quantity
${\cal H}\equiv {a}'/a$ is the conformal Hubble parameter defining the expansion rate of the Universe. Its present value $H_0$ calls the Hubble constant. The energy continuity equation in conformal time
\be\label{continuityclassical}
{\rho}'=-3(\rho+P)\left(\frac{{a}'}{a}\right)
\ee
with an equation of state $P=w\rho$, connecting the pressure $P$ and density $\rho$, yields the matter density sources dependence of the scale factor
\be
\rho=\rho_0\left(\frac{a_0}{a}\right)^{3(1+w)}.
\ee
Here, $\rho_0$ and $a_0$ are the modern values of the corresponding characteristics.
So, for simplest cases, one has

$\bullet$
interstellar dust: $w=0,$ $P=0: \rho =\rho_0 \left({a_0}/{a}\right)^{3};$

$\bullet$
radiation: $w=1/3$, $P=\rho/3: \rho =\rho_0\left({a_0}/{a}\right)^{4};$

$\bullet$
De Sitter vacuum: $w=-1$, $P=-\rho: \rho=\rho_0.$

Taking into consideration the vacuum density, non-relativistic and relativistic matter with the corresponding cosmological parameters $\Omega_{\rm \Lambda}$, $\Omega_{\rm M}$, $\Omega_{\rm rad},$ we get the sum of their contributions
\be\label{rho}
\rho=\frac{3H_0^2}{8\pi G}
\left[
\Omega_{\rm\Lambda}+\Omega_{\rm M}\left(\frac{a_0}{a}\right)^3+
\Omega_{\rm rad}\left(\frac{a_0}{a}\right)^4
\right].
\ee
The cosmological parameters are constrained
\begin{equation}\label{constraint}
\Omega_{\rm\Lambda}+\Omega_{\rm M}+\Omega_{\rm rad}+\Omega_{\rm curv}=1,
\end{equation}
where for curvature term is
\begin{equation}\label{curv}
\Omega_{\rm curv}\equiv -\frac{k}{a_0^2H_0^2}.
\end{equation}
Substituting (\ref{rho}), (\ref{constraint}), (\ref{curv}) into the right hand side of the conformal Friedmann equation (\ref{Friedmann}), we obtain
\bea\nonumber
&&\frac{1}{x^2}\left(\frac{{\rm d}x}{{\rm d}\eta}\right)^2=\\
&=&\left(\frac{H_0 a_0}{c}\right)^2\left[\Omega_{\rm \Lambda}x^2+\Omega_{\rm curv}+\frac{\Omega_{\rm M}}{x}+
\frac{\Omega_{\rm rad}}{x^2}\right].\label{Friedmannclass}
\eea
Here the variable $x$ is given as a ratio of a scale $a(\eta)$ to a modern value $a_0$:
\begin{equation}\label{xequiv}
x\equiv\frac{a(\eta)}{a_0}=\frac{1}{1+z},
\end{equation}
$z$ is a redshift of spectral lines,
\be\label{Hubbleconstant}
H_0=h\cdot 10^5 m/s/Mpc,\quad h=0.72\pm 0.08
\ee
is the Hubble constant.

From the above differential equation (\ref{Friedmannclass}) the formula for getting the conformal time as a function of the scale is followed:
\bea
\eta(x)&=&\frac{c}{H_0 a_0\sqrt{\Omega_\Lambda}}\int\limits_0^x\frac{{\rm d}x}
{\sqrt{x^4+ 6a_2 x^2+4a_3 x+a_4}}\nonumber\\
&\equiv&
\frac{c}{H_0 a_0\sqrt{\Omega_\Lambda}}I(x).
\label{auxiliary}
\eea
Here the following notations are introduced
\be\nonumber
6a_2\equiv\left(\frac{\Omega_{\rm curv}}{\Omega_{\rm\Lambda}}\right),\quad
4a_3\equiv\left(\frac{\Omega_{\rm M}}{\Omega_{\rm\Lambda}}\right),\quad
a_4\equiv\left(\frac{\Omega_{\rm rad}}{\Omega_{\rm\Lambda}}\right).
\ee
Notice, that it is possible to consider the subradical polynomial of generic kind
\be\label{P(x)}
P(x)=x^4+4A_1 x^3+6A_2 x^2+4A_3 x+A_4.
\ee
The second term $4A_1 x^3$ corresponds to domain walls contribution~\cite{Fil} characterized by

$\bullet$
$w=-2/3,$ $p=-(2/3)\rho: \rho=\rho_0(a_0/a).$

However, by shift the variable $x\mapsto x-A_1$ the polynomial $P(x)$ (\ref{P(x)}) is reduced to the standard form
\be\label{P4}
P_4(x)\equiv x^4+ 6a_2 x^2+4a_3 x+a_4
\ee
with the coefficients
\be\nonumber
a_2=A_2-A_1^2,\quad a_3=A_3-3A_1A_2+2A_1^3,
\ee
\be\nonumber
a_4=A_4-4A_1A_3+6A_2A_1^2-3A_1^4.
\ee

The polynomial (\ref{P4}) is of the fourth order, to low its order let us implement the change of the variable $x\mapsto y$ by the rule~\cite{PavlovMIPh}:
\be\label{subs}
\sqrt{x^4+ 6a_2 x^2+4a_3 x+a_4}\equiv x^2-2y+a_2.
\ee
Raising to square the both sides of the equation (\ref{subs}), one gets the polynomial expression
\be\label{polynomial}
4a_2 x^2+4x^2 y+4 a_3 x-4 y^2+4 a_2 y+a_4-a_2^2=0.
\ee
The differential of the equality (\ref{polynomial}) can be written in the form of a proportion
\be\label{proportion}
\frac{{\rm d}x}{x^2-2y+a_2}=-\frac{{\rm d}y}{2a_2 x+2xy+a_3}.
\ee
With regard to equality (\ref{subs}), we rewrite the proportion (\ref{proportion}) in the form
\be\label{pro}
\frac{{\rm d}x}{\sqrt{x^4+ 6a_2 x^2+4a_3 x+a_4}}=-\frac{{\rm d}y}{2a_2 x+2xy+a_3}.
\ee
Let us express the variable $x$ from the (\ref{polynomial})
\be\label{x}
x(y)=\frac{-a_3\pm\sqrt{4y^3-g_2 y-g_3}}{2(y+a_2)},
\ee
where the following notations were introduced:
\be\nonumber
g_2\equiv a_4+3a_2^2,\qquad g_3\equiv a_2a_4-a_3^2-a_2^3.
\ee
Let us express the variable $y$ from the equality (\ref{polynomial})
\be\label{subsy(x)}
y(x)=\frac{1}{2}\left[x^2+a_2\pm\sqrt{x^4+6a_2 x^2+4 a_3 x+a_4}\right].
\ee

The equality (\ref{x}) can be presented in the form
\be
\left(2xy+2a_2 x+a_3\right)^2=4y^3-g_2 y-g_3.
\ee
Then we implement the next substitution $y\mapsto u$:
\be\label{ywp}
y=\wp (u;g_2,g_3),
\ee
where $\wp (u;g_2,g_3)$ is the Weierstrass function ~\cite{Whittaker}.
Then the differential equation (\ref{pro}) takes the exact form
\be
\frac{{\rm d}y}{2a_2 x+2xy+a_3}=\frac{{\rm d}y}{\sqrt{4y^3-g_2 y-g_3}}={\rm d}u.
\ee

The integral (\ref{auxiliary}) can be presented to the standard form after the substitution (\ref{subsy(x)})
\bea
&&I(x)\equiv\int\limits_0^x\frac{{\rm d}x}{\sqrt{P_4(x)}}=-
\int\limits_{(a_2-\sqrt{a_4})/2}^{y(x)}\frac{{\rm d}y}{\sqrt{4y^3-g_2 y-g_3}}\nonumber\\
&&-\int\limits_{(a_2-\sqrt{a_4})/2}^{+\infty}\frac{{\rm d}y}{\sqrt{4y^3-g_2 y-g_3}}
+\int\limits_{y(x)}^{+\infty}\frac{{\rm d}y}{\sqrt{4y^3-g_2 y-g_3}}.\nonumber
\eea
Now, after the substitution (\ref{ywp}), the integral is expressed via the inverse Weierstrass functions
\bea\nonumber
I(x)&=&-\int\limits_{\wp^{-1}[(a_2-\sqrt{a_4})/2]}^{+\infty}{\rm d}u+
\int\limits_{\wp^{-1}[y(x)]}^{+\infty}{\rm d}u\\
&=&-\wp^{-1}\left[(a_2-\sqrt{a_4})/2)\right]+\wp^{-1}[y(x)].\label{inv}
\eea
Finally, we yield the exact solution: the conformal time as a function of the scale $\eta=\eta (x)$:
\be\label{eta(x)}
\frac{H_0a_0\sqrt{\Omega_{\rm\Lambda}}}{c}\eta (x)=-\wp^{-1}\left[(a_2-\sqrt{a_4})/2\right]
+\wp^{-1}\left[y(x)\right].
\ee
The expression (\ref{eta(x)}) can be converted
\be\nonumber
\wp [u]=y(x),
\ee
where we denoted
\be\nonumber
u\equiv
\frac{H_0 a_0\sqrt{\Omega_{\rm\Lambda}}}{c}\eta (x)+\wp^{-1}[y(0)].
\ee

\section{Periods of doubly periodic functions}

Let us focus our attention on the cubic polynomial in Weierstrass form
\be\label{polynomialstandard}
P_3(y)\equiv 4y^3-g_2 y-g_3,
\ee
where $g_2, g_3$ are its invariants.
Depending on the sign of the discriminant of the polynomial, we have three cases, which we consider in order.

$\bullet$
The discriminant
\be\label{discriminant}
\Delta\equiv g_2^3-27g_3^2
\ee
is positive: $\Delta>0$.
Then the roots of the polynomial (\ref{polynomialstandard}) are real and also $e_1>e_2>e_3$:
\be\nonumber
e_2=s_1+s_2,
\ee
where
\bea\nonumber
&&s_1\equiv\frac{1}{2}\left[g_3+\frac{1}{3\sqrt{3}}\sqrt{-\Delta}\right]^{1/3},\\
&&s_2\equiv\frac{1}{2}\left[g_3-\frac{1}{3\sqrt{3}}\sqrt{-\Delta}\right]^{1/3};\nonumber
\eea
and the rest two are:
\bea\nonumber
&&e_1=-\frac{1}{2}(s_1+s_2)+\frac{\imath\sqrt{3}}{2}(s_1-s_2);\\
&&e_3=-\frac{1}{2}(s_1+s_2)-\frac{\imath\sqrt{3}}{2}(s_1-s_2).\nonumber
\eea

The $\wp$-function has one real semi-period $\omega_1$
\be\nonumber
\omega_1=\int\limits_{e_1}^{+\infty}\frac{{\rm d}y}{\sqrt{4(y-e_1)(y-e_2)(y-e_3)}},
\ee
and one is imagine $\omega_3$
\be\nonumber
\omega_3=\imath\int\limits_{-\infty}^{e_3}\frac{{\rm d}y}{\sqrt{4(e_1-y)(e_2-y)(e_3-y)}}.
\ee

The Weierstrass function is two-parameter, and for practical use it is more convenient to work with the one-parameter elliptic Jacobi function.
If we change the variable
\be\nonumber
\wp (u)=e_3+\frac{e_1-e_3}{{\rm sn}^2 (u\sqrt{e_1-e_3})}
\ee
with the modulus of the elliptic sine
\be\nonumber
k=\sqrt{\frac{e_2-e_3}{e_1-e_3}},
\ee
we yield the solution (\ref{eta(x)}) in Jacobi form
\be
{\rm sn}^2 (u\sqrt{e_1-e_3})=\frac{e_1-e_3}{y(x)-e_3}.
\ee
The basic periods of the elliptic sine are $4K$ and $2K'$, where
\be\nonumber
K\equiv\int\limits_0^{\pi/2}\frac{{\rm d}\phi}{\sqrt{1-k^2\sin^2\phi}},\quad
K'\equiv\int\limits_0^{\pi/2}\frac{{\rm d}\phi}{\sqrt{1-k'^2\sin^2\phi}},
\ee
and $k^2+k'^2=1$.

$\bullet$
The discriminant (\ref{discriminant}) is negative: $\Delta<0$. Therefore, one root $e_2$ of the polynomial (\ref{polynomialstandard}) is real:
\be\nonumber
e_2=s_1+s_2,
\ee
and the rest two are complex conjugated:
\be\nonumber
e_1= m+\imath n;\qquad e_3= m-\imath n.
\ee
Present the polynomial (\ref{polynomialstandard}) as a product
\be\nonumber
4y^3-g_2 y-g_3=4(y-e_2)[(y-m)^2+n^2].
\ee
The semi-periods of the Weierstrass $\wp$-function:
\be\nonumber
\omega_2=\int\limits_{e_2}^{+\infty}\frac{{\rm d}y}{\sqrt{4(y-e_2)[(y-m)^2+n^2]}},
\ee
\be\nonumber
\omega_2'=\imath\int\limits_{-\infty}^{e_2}\frac{{\rm d}y}{\sqrt{4(e_2-y)[(y-m)^2+n^2]}}.
\ee

If we make the substitution
\be
\wp (u)=e_2+H\frac{1+{\rm cn} (2u\sqrt{H})}{1-{\rm cn} (2u\sqrt{H})}
\ee
with $H=\sqrt{9m^2+n^2}$ we present the solution (\ref{eta(x)}) in Jacobi form
\be\nonumber
{\rm cn}(2u\sqrt{H})=
\frac{y(x)-(e_2+H)}{y(x)-(e_2-H)}.
\ee
where the modulus is equal to
\be\nonumber
k\equiv\sqrt{\frac{1}{2}-\frac{3e_2}{4H}}.
\ee
Basic periods for the elliptic cosine are
$4K$ and $2K+2\imath K'$.

$\bullet$
The discriminant (\ref{discriminant}) is equal to zero: $\Delta=0$.
Therefore, all roots of the polynomial (\ref{polynomialstandard}) are real. If invariants $g_2$ and $g_3$ are not zero, two roots are equal.
Let $e_1=e_2$, then $k^2\equiv (e_2-e_3)/(e_1-e_3)=1$.
Weierstrass function $\wp (u)$ is expressed via hyperbolic one.

Let $e_2=e_3$, then $k^2\equiv (e_2-e_3)/(e_1-e_3)=0$.
Weierstrass function $\wp (u)$ is expressed via trigonometric sine:
\be\nonumber
\wp (u)=-\frac{3g_3}{2g_2}+\frac{9g_3}{2g_2}\frac{1}{{\rm sin}^2\left(u\sqrt{\frac{9g_3}{2g_2}}\right)}.
\ee

If invariants $g_2$ and $g_3$ are zero, all three roots are equal.
Then, one yields
\be\nonumber
u=\int\limits_y^{+\infty}\frac{{\rm d}y}{2\sqrt{y^3}}=\frac{1}{\sqrt{y}},
\ee
and the substitution is rational
\be\nonumber
y=\wp (u)=\frac{1}{u^2}.
\ee

\section{Friedmann equation in coordinate time}

In the standard cosmology the Friedmann equation in coordinate time~\cite{Weinberg}:
\begin{equation}\label{Friedmannstandard}
\left(\frac{\dot{a}}{a}\right)^2=\frac{8\pi G}{3}\rho-\frac{k}{a^2},
\end{equation}
where the dot denotes differentiation with respect to coordinate time $t$.
By analogy with the quadrature in conformal time (\ref{auxiliary}) we get the quadrature in coordinate time
\bea
t&=&\frac{c}{H_0\sqrt{\Omega_\Lambda}}\int\limits_0^x\frac{x{\rm d}x}
{\sqrt{x^4+ 6a_2 x^2+4a_3 x+a_4}}\nonumber\\
&=&\frac{c}{H_0\sqrt{\Omega_\Lambda}}\int\limits_0^x\frac{x{\rm d}x}{\sqrt{P_4(x)}}
\equiv
\frac{c}{H_0\sqrt{\Omega_\Lambda}}J(x).
\label{auxiliarystandard}
\eea
After substitution  $x\mapsto y$ (\ref{x}) one gets for the integral the expression
\be\label{third}
J(x)=\int\limits_{(a_2-\sqrt{a_4})/2}^{y(x)}\frac{x(y){\rm d}y}{\sqrt{P_3(y)}},
\ee
with the polynomial of the third order (\ref{polynomialstandard}).
For the variable $x(y)$ in (\ref{third}) we have the substitution (\ref{x}) in terms of Weierstrass function~(\ref{ywp}):
\bea\nonumber
x[y(u)]&=&\frac{-a_3\pm\sqrt{4\wp^3(u)-g_2\wp(u)-g_3}}{2(\wp(u)+a_2)}\\
&=&\frac{-a_3+\wp'(u)}{2(\wp (u)+a_2)}.\label{xwpsubs}
\eea
After substitution (\ref{xwpsubs}) into (\ref{third}), the integral (\ref{third}) takes the form
\bea\nonumber
&&J(x)=-\frac{a_3}{2}\int\limits_{\wp^{-1}[(a_2-\sqrt{a_4})/2]}^{\wp^{-1}[y(x)]}\frac{{\rm d}u}{\wp (u)+a_2}\\
&&+\frac{1}{2}\int\limits_{\wp^{-1}[(a_2-\sqrt{a_4})/2]}^{\wp^{-1}[y(x)]}\frac{\wp'(u){\rm d}u}{\wp (u)+a_2}\equiv J_1+J_2.\label{indefinite}
\eea
The second integral is equal to
\be\label{J2}
J_2=\frac{1}{2}\ln\left(\frac{y(x)+a_2}{(a_2-\sqrt{a_4})/2+a_2}\right).
\ee
If in the first integral $a_2$ is not equal to the semi-period of the function $\wp (u)$, denoting
\be
a_2\equiv -\wp (v),
\ee
we have for the underintegrand
\be\label{J1}
\frac{1}{\wp (u)-\wp (v)}=-\frac{1}{\wp'(v)}\left[\zeta (u+v)-\zeta (u-v)-2\zeta (v)\right].
\ee
Here we introduced the Weierstrass function
\be\label{zeta}
\zeta' (u)=-\wp (u),
\ee
Integration yields
\bea\nonumber
&&J_1=-\frac{a_3}{2}
\int\limits_{\wp^{-1}[(a_2-\sqrt{a_4})/2]}^{\wp^{-1}[y(x)]}
\frac{{\rm d}u}{\wp (u)-\wp (v)}\\
&&=-\frac{a_3}{2\wp' (v)}\ln\left(\frac{\sigma (\wp^{-1}[y(x)]-v)}{\sigma (\wp^{-1}[y(x)]+v)}\right)\nonumber\\
&&+\frac{a_3}{2\wp' (v)}\ln\left(\frac{\sigma (\wp^{-1}[(a_2-\sqrt{a_4})/2]-v)}
{\sigma (\wp^{-1}[(a_2-\sqrt{a_4})/2]+v)}\right)\nonumber\\
&&-\frac{a_3\zeta (v)}{\wp' (v)}\left(\wp^{-1}[y(x)]-\wp^{-1}[(a_2-\sqrt{a_4})/2]\right)
.\label{intJ1}
\eea
Here we introduced once more Weierstrass function
\be\label{sigma}
\left(\ln \sigma (u)\right)'=\frac{\sigma'(u)}{\sigma (u)}=\zeta (u).
\ee
The obtained solutions of the Friedmann equation in coordinate time look rather complicated.
Therefore, it is more convenient to work with the Friedmann conformal equation.

\section{Discussion}

In theoretical mechanics, the solution of a mathematical pendulum problem with length of thread $l$ is expressed in terms of an elliptic sine. Two periods of the doubly periodic Jacobi function reveal dynamical sense.
The period of oscillations is equal to
\be\nonumber
T=4\sqrt{\frac{l}{g}}K,
\ee
where $g$ is the acceleration of gravity. $K$ is the complete elliptic integral of the first kind
\be\nonumber
K=\int\limits_0^1\frac{{\rm d}t}{\sqrt{(1-t^2)(1-k^2t^2)}},
\ee
with modulus $k^2=h/(2l)$, and $h$ is an initial hight of the point.
Paul Appel noticed ~\cite{Whitt} that the imaginary period is equal to
\be\nonumber
T'=4\sqrt{\frac{l}{g}}K'
\ee
where $K'$ is the additional elliptic integral with
\be\nonumber
k'^2=1-k^2=1-\frac{h}{2l}=\frac{(2l-h)}{2l}.
\ee
It corresponds to the pendulum whose initial hight is equal to $(2l-h)$.

According to Roger Penrose ideas of conformal cyclic cosmology the universe undergoes cycles during its evolution \cite{Penrose}. The history of the Universe is considered without an inflation stage.
One eon is a continuation of an other one. There was eon antecedent to the Big Bang.
So, it should be another eon after the Big Bang. This picture is prolonged to both directions: in past and towards future. The conformal cyclic cosmology predicts the presence of families of concentric low-variance circular rings in the cosmic microwave background picture~\cite{Gurzadyan}.

In quantum cosmology, the wave function of the universe is studied in problems of its quantum origin~\cite{Vilenkin}.
Quantum tunnel transitions with a change in the signature of spacetime are described in the imaginary time
formalism~\cite{Alt}. The description of the classically forbidden state using imaginary time means the complexification of the conformal superspace~\cite{PavlovExtrinsic}.
This complexification makes the conformal time variable purely imaginary and transforms the Wheeler--DeWitt equation from hyperbolic to elliptic. This is analogous to the transition from the hyperbolic to the elliptic Klein--Gordon equation for Wick rotation.

It is possible to consider exotic states of medium with EoS $P=w\rho$ for various fractional values $w$ with integer $n=3(1+w)$~\cite{Fil}. The problem of integration of the Friedmann equation is reduced to Abel integrals
for algebraic functions~\cite{Golubev}. The Abel integrals for algebraic functions are named hyperelliptic.
In this work there was found the class of solutions of the Friedmann equation. It was proven that the solutions belong to the Weierstrass doubly-periodic functions. It can take an addition information for understanding the evolution of the Universe.



\begin{thebibliography}{10}

\bibitem{RiessNobel}
A.~G.~Riess,
{Rev. Mod. Phys.} {\bf 84}, 1165 (2012).

\bibitem{Behnke}
D.~Behnke, et al., Phys. Lett. {\bf B 530}, 20 (2002).

\bibitem{Zakharov:2010nf}
A.~F.~Zakharov, V.~N.~Pervushin,
{Int. J. Mod. Phys. D} {\bf 19}, 1875 (2010).

\bibitem{PavlovMIPh}
  A.~E.~Pavlov,
  {RUDN J. Math. Inform. Sc. Phys.} {\bf 25}, 390 (2017).

\bibitem{Golubev}
V.~V.~Golubev,
{\it Lectures on Integration of the Equations of Motion of a Rigid Body about a Fixed Point} (URSS, Moscow, 2021).

\bibitem{Gerasimov}
I.~A.~Gerasimov, {\it Weierstrass Functions and their Applications to Mechanics and Astronomy}
(Moscow State University Press, Moscow, 1990).

\bibitem{PenroseRus}
R.~Penrose, Hypercomplex Numbers in Geometry and Physics. {\bf 10}, 62 (2013).

\bibitem{Weinberg}
S.~Weinberg, {\it Cosmology} (Oxford University Press, Oxford, 2008).

\bibitem{Fil}
M.~L. Filchenkov, S.~V.~ Kopylov, V.~S.~ Evdokimov, {\it Gravitation, Astrophysics, Cosmology}
(URSS, Moscow, 2017).

\bibitem{Whittaker}
E.~T.~Whittaker, G.~N.~Watson, {\it A Course of Modern Analysis} (Cambridge University Press, Cambridge, 1927).

\bibitem{Whitt}
E.~T.~Whittaker, {\it A Treatise on the Analytical Dynamics of Particles and Rigid Bodies}
(Cambridge University Press, Cambridge, 1927).

\bibitem{Penrose}
R.~Penrose, {\it Cycles of Time: An Extraordinary New View of the Universe} (The Bodley Head, London, 2010).

\bibitem{Gurzadyan}
V.~G.~Gurzadyan, ~R.~Penrose, Eur. Phys. J. Plus. {\bf 128}, 22 (2013).

\bibitem{Vilenkin}
A.~Vilenkin, Phys. Rev. {\bf D 33}, 3560 (1986).

\bibitem{Alt}
B.~L. ~Al'tshuler, A.~O.~ Barvinsky, Physics-Uspekhi. {\bf 39}, 429 (1996).

\bibitem{PavlovExtrinsic}
A.~E.~ Pavlov, Grav. Cosmol. {\bf 26}, 208 (2020).

\end{thebibliography}
\end{document}